# Introducing A Public Stereoscopic 3D High Dynamic Range (SHDR) Video Database

Amin Banitalebi-Dehkordi
University of British Columbia (UBC), Vancouver, BC, Canada
dehkordi@ece.ubc.ca

*Abstract*—**High Dynamic Range (HDR) displays and cameras are paving their ways through the consumer market at a rapid growth rate. Thanks to TV and camera manufacturers, HDR systems are now becoming available commercially to end users. This is taking place only a few years after the blooming of 3D video technologies. MPEG/ITU are also actively working towards the standardization of these technologies. However, preliminary research efforts in these video technologies are hammered by the lack of sufficient experimental data. In this paper, we introduce a Stereoscopic 3D HDR (SHDR) database of videos that is made publicly available to the research community. We explain the procedure taken to capture, calibrate, and post-process the videos. In addition, we provide insights on potential use-cases, challenges, and research opportunities, implied by the combination of higher dynamic range of the HDR aspect, and depth impression of the 3D aspect.**

I. INTRODUCTION

Over the past few years industry and research communities have been showing an increasing amount of interest in immersive media technologies. These technologies aim at providing a higher, closer to natural, Quality of Experience (QoE) to end users. Examples of these technologies include 3D video, High Dynamic Range (HDR) video, 360º displays, Virtual Reality (VR), and 4K to 8K video. MPEG and ITU are also holding active discussions in regards to the standardization of such types of media content [1]. These efforts received a boost after the introduction of the first draft of HEVC (High Efficiency Video Coding) [29,30].

Considering that some of these technologies are fairly new, free and public availability of these databases is still a challenge. This paper focuses on two specific state-of-the-art video technologies namely 3D and HDR video.

3D video technologies entered the consumer market a few years back, by introducing stereoscopic video. Stereo video contains two left and right views, which are calibrated and aligned in such a way that provide an impression of depth (through disparity) to viewers. Initial commercialization efforts in the realm of 3D technology included the standardization of stereo and multi-view video. Later many showed interest in glass-free auto-stereo TVs and several models became commercially available. Since then, many 3D movies were produced and displayed at theaters. However, there are still only limited number of public databases of original uncompressed 3D video available. Note that 3D movies produced for cinema, are almost all the time converted from 2D. In other words, the movie is initially captured in 2D, but then later 2D shots are manually converted to 3D in post-production studios. That being said, compared to other kinds of immersive video technologies mentioned earlier, there are more 3D databases (mainly stereo video) available [2-3, 25]. Consequently, research in areas of 3D video compression [26-28,1,4,35-36], quality assessment [31-33,5-8], and 3D visual attention modeling [34,9-10] is very active.

High Dynamic Range video technologies have been receiving significant recognition over the last couple of years. HDR video provides a wide range of dynamic range, through a close to Human Visual System (HVS) perceived gamut. As opposed to Standard Dynamic Range (SDR) video that spans over only a contrast range of 100:1 to 1000:1, HDR video can potentially cover up to $10^5$:1 contrast ratio [11]. Each pixel in HDR video frame stores more than 8 bits of information, often up to 16 bits. This results in ordinary SDR displays not being able to show HDR content. To show HDR content on regular displays a technique known as tone mapping is employed to reduce the dynamic range to 8 bits, in a way that most of the color and brightness information is preserved [12]. In addition, capturing HDR video requires special camera technologies that support capturing of many exposure levels simultaneously [11, 12]. This make it harder to capture and produce freely available HDR databases. As a result there are only very few HDR databases publicly available. As well, research on HDR video compression [13,14, 37-39], HDR quality assessment [15,40,41], and HDR visual attention modeling [16] are at preliminary stages.

One of advanced video technologies that is recently being discussed within the research and industry communities is the combination of 3D and HDR video, known as 3D-HDR video. Stereoscopic HDR (SHDR) is a two-view video with each view being captured in HDR. It is worth noting that early efforts in combining 3D and HDR content was shaped around creating synthetic HDR images using multiple views of a same scene [21-22]. [23] proposes to use side-by-side inexpensive low dynamic range cameras to generate HDR stereo data after further processing of the two. [24] takes this one step ahead by proposing to use one LDR and one HDR camera to generate stereo HDR data, that was shown in [24] to be very closely similar to the ground truth captured stereo HDR content. However, it was only recently that HDR video cameras became commercially available (and affordable). As a result, none of the mentioned studies used multiple HDR cameras to capture stereo or multi view HDR content. As 3D-HDR video receives more attention, it would be necessary to have standard public databases for performing research activities. To the best of the author's knowledge, these is no public large-scale database of SHDR video available to this date.

This paper introduces a public database of SHDR videos. Stereoscopic videos are captured using two side by side HDR cameras. After that, a process of calibration and rectification is performed to align the videos to ensure a pleasant 3D quality is achieved. Our SHDR video database is available at: http://ece.ubc.ca/~dehkordi/databases.html

The proposed database contains 11 videos from a wide range of brightness, depth, color, temporal, and spatial complexity. The videos are captured and calibrated accurately both at the hardware (physically locking and aligning cameras) and post-processing (disparity correction and frame alignment) levels.

The main contribution of this paper is the introduction of a new 3D-HDR database to the community, to facilitate the investigation of challenges involved with processing and understanding this kind of content. In addition, the creation of the database is explained in reproducible steps so the same approach can be carried out by colleagues to generate new 3D-HDR content from different scenes with different capturing or lighting settings, or with other cameras. It is worth mentioning that this paper does not attempt to solve challenges involved with processing or understanding of 3D-HDR data, but rather it only re-iterates the primary challenges and potential areas of research for interested readers.

Section II introduces our database and explains the procedure to capture and post-process the videos. Section III elaborates on the use cases and challenges involved in 3D-HDR video processing, that exist solely in 3D-HDR video. Section IV concludes the paper.

## II. Preparation of our 3D-HDR video database

This section provides a description on the capturing setup, database scene content, as well as post-processing stages involved with the content creation.

### A. Capturing Configuration

The two RED Scarlet-X cameras used for capturing are mounted in a side-by-side fashion on top of a tripod. As Fig. 1 illustrates, we use an adjustable metal bar to mount the cameras that can support up to 4 side-by-side cameras. The bar, joints, and screws are built to a sub-millimeter precision. Since we only capture two view (stereo) HDR video, we use only two stands on the bar (multi-view HDR video can be captured similarly). The cameras use identical firmware and settings, and are of a same model and build date. A single remote control is used to control both cameras. The two cameras were synchronized using a video Genlock input signal (RS170A Tri-Level Sync Input) [47]. There was about 8 cm horizontal disparity between the centers of the lenses. Each camera records videos at 30 fps and in floating points HDR (18 F-stops). For indoor sequences an artificial light source was used to provide increased dynamic range and brightness over the target objects.

## B. Capturing Content

Using our capturing setup described in the previous sub-section, over 24 sequences were captured initially. All videos are shot in full HD resolution at 1920×1080 (each view) and 30 fps. From the initially captured sequences, a total of 11 videos were selected for our database, to cover different ranges of scene content, lighting conditions, and pleasant 3D quality. Fig. 2 demonstrates a snapshot of the videos and Table 1 provides specifications on the sequences. Note that the snapshots in Fig. 2 are only from one view of each video. Also, in order to be able to show the snapshots here, we converted floating point frames to 8 bits through tone-mapping [12].

Generally the selected videos provide an impression of depth, as well as an impression of a higher dynamic range, that comes with not just a brighter video, but also more information at different exposures. Each video has about 10 second length which makes it possible to use this database for various types of subjective experiments.

As it is observed from Table 1, videos cover a different ranges of spatial and temporal complexity. Spatial complexity, measured as Spatial Information (SI), is a measure of spatial complexity of a scene [17]. SI is calculated by applying a Sobel filter to each frame (luminance plane) to detect edges first. Then standard deviation on the edge map provides a measure of spatial complexity of the associated frame. Maximum standard deviation value over all frames results in the SI value of the sequence. Temporal Information (TI) on the other hand, measures temporal complexity, and is based on motion among consecutive frames [17]. To measure the TI, first the difference between the pixel values (of the luminance plane) at the same coordinates in consecutive frames is calculated. Then, the standard deviation over pixels in each frame is computed and the maximum value over all the frames is set as the measure of TI. Note that SI and TI were calculated over the entire dynamic range of the SHDR videos, after they were converted to 12 bits (more details later in this Section). Therefore, the SI and TI metrics measure the complexity in spatial and temporal domain in the brightness intensity units as digital numbers and thus can be considered unit-less. Fig. 3 shows the SI and TI distribution over the proposed video database.

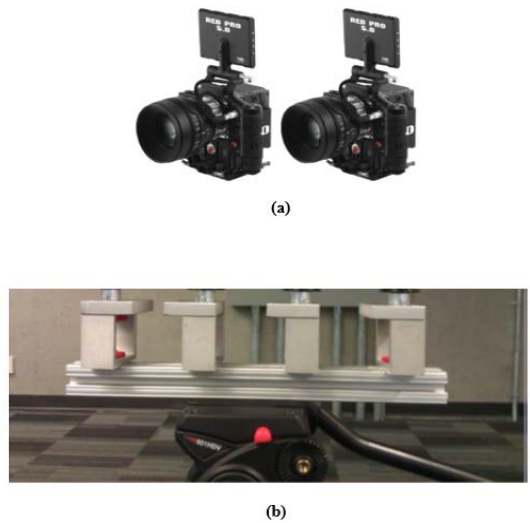

(a)

(b)

**Fig. 1. Capturing with two side-by-side HDR cameras. Cameras are mounted to an adjustable aluminum bar on a tripod**

Table 1 also provides information about "depth bracket" of each scene. The depth bracket for each scene is defined as the amount of 3D space used in a shot (or a sequence). In other words, it is a rough estimate of difference between the distance of the closest and the farthest visually important objects from the camera in each scene [45].

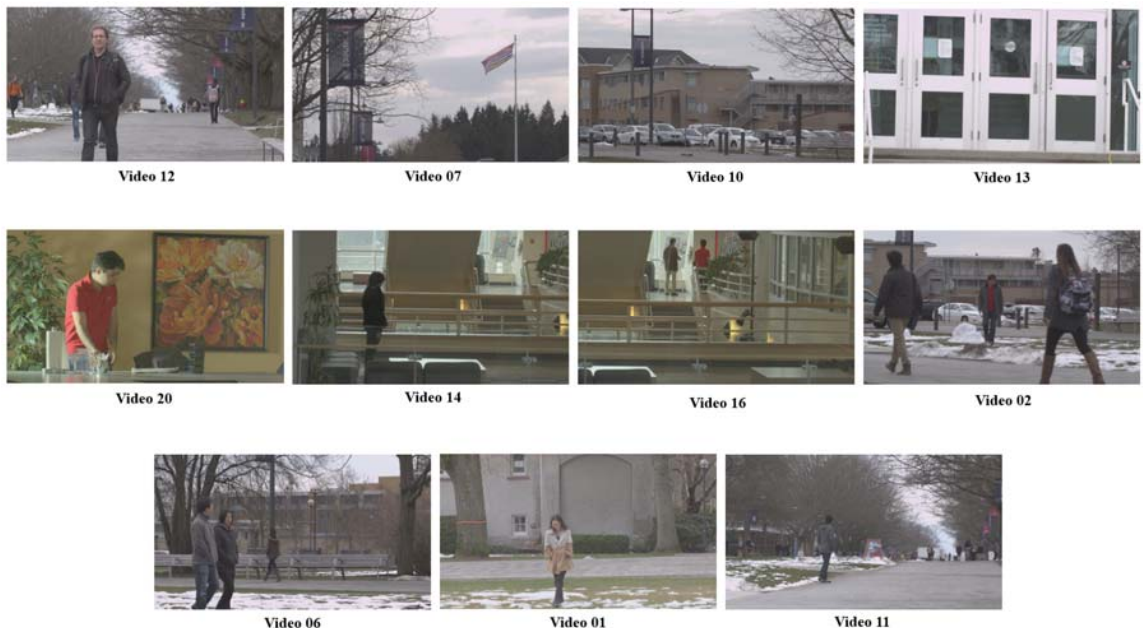

Fig. 2. Snapshots of tone-mapped video frames. Only right view frames are shown here.

Table 1 Specifications of our 3D-HDR video database

| Sequence | Resolution | Frame Rate (fps) | Number of Frames | Spatial Complexity (Spatial Information) | Temporal Complexity (Temporal Information) | Depth Bracket |
|---|---|---|---|---|---|---|
| **Video_01** | 1920×1080 | 30 | 300 | Low (26.8121) | Low (5.8977) | Medium |
| **Video_02** | 1920×1080 | 30 | 510 | Medium (32.5871) | High (26.5389) | Wide |
| **Video_06** | 1920×1080 | 30 | 300 | High (49.3520) | High (20.8599) | Wide |
| **Video_07** | 1920×1080 | 30 | 300 | High (50.7216) | Low (5.3762) | Wide |
| **Video_10** | 1920×1080 | 30 | 460 | Medium (34.6945) | Low (2.4159) | Wide |
| **Video_11** | 1920×1080 | 30 | 440 | High (45.0262) | Low (7.5869) | Wide |
| **Video_12** | 1920×1080 | 30 | 300 | Medium (35.9981) | Medium (12.4612) | Wide |
| **Video_13** | 1920×1080 | 30 | 400 | Medium (39.1949) | Medium (11.2539) | Medium |
| **Video_14** | 1920×1080 | 30 | 400 | Low (25.5698) | Low (5.5410) | Medium |
| **Video_16** | 1920×1080 | 30 | 300 | Low (23.3037) | Medium (9.7458) | Medium |
| **Video_20** | 1920×1080 | 30 | 170 | Low (23.3551) | Medium (10.6995) | Narrow |

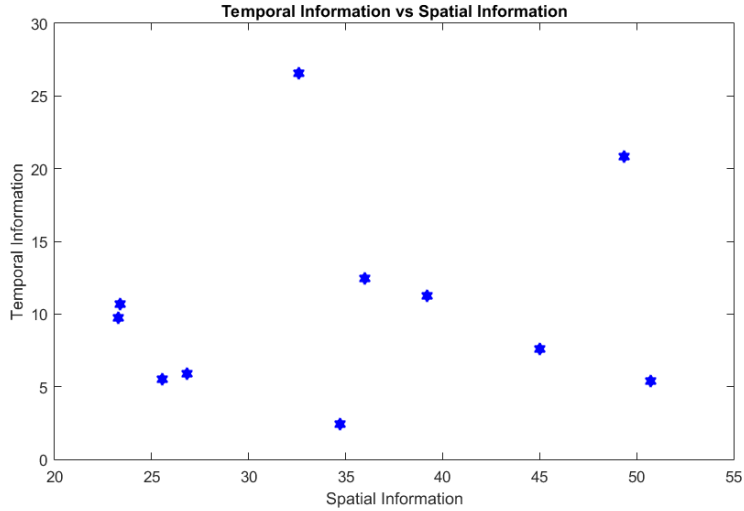
**Fig. 3. Spatial and temporal information associated to the SHDR video database.**

When capturing, we tried to avoid any 3D window violation. Window violation (in 3D) occurs when objects only partially appear in a scene. As a result, poor 3D quality (due to conflict between depth cues) is expected around the edges of the 3D display.

The captured sequences are first stored in ".hdr", floating point format, and in linear RGB space. However, to make these videos useful for 3D and HDR video compression studies, they were converted to 12 bit "4:2:0" raw format according to ITU-R Rec. BT.709 color conversion primaries [42] (the HEVC and H.264/AVC video encoders currently only support the compression of video in ".yuv" raw file format). To this end, the RGB values were first normalized to [0, 1]. Then, they were converted to Y-Cb-Cr color space in accordance to ITU-R Rec. BT.709 [42]. At last, using the separable filter values provided by [43], chroma-subsampling was applied to the Y-Cb-Cr values to linearly quantize the resulted signals to the desired bit depth (i.e., 12 bits) [44]. Both the raw ".hdr" stored format, and the ".yuv" format will be made publicly available along with this paper.

*C. Post-Processing*

Several stages of post-processing were applied to the captured sequences to ensure they have a high 3D and HDR quality and are comfortable to watch.

*1) Temporal Synchronization*

Although a same remote control was used, and camera firmware versions were exactly the same, but due to very small hardware differences, it is always possible that the two left and right cameras are out of sync, even for a couple of frames. This makes more sense considering that video frame rate for each view is 30 fps, so even if there is 1/30$^{th}$ of a second difference, that can cause temporal mismatch of one frame.

Two ensure temporal synchronization, we manually marked objects of interest, and compared them in the two views. Wherever needed, a few frames were taken out to make sure the views are perfectly synchronized.

*2) 3D Alignment*

It was mentioned previously that our camera configuration hardware was built up to a sub-millimeter precision. However, since cameras are manually mounted (and taken apart after), there is a small possibility that there is very minor vertical misalignment in the configuration. Any vertical mismatch can make the 3D content uncomfortable to watch.

To remove any vertical mismatch, we extract SIFT [18] features of the two views, and match them against each other. Top 10 % of the matches are selected as reliable matches, and average height difference between those is an amount that is cropped from one of the views. This provides vertical alignment and stability of frames [19].

*3) Disparity Correction*

Using two side-by-side cameras results in negative parallax as objects converge in infinity (objects pop out of the display). This can cause visual discomfort to the viewers as subjective experiments have shown viewers show preference in watching objects appearing behind the display (positive parallax), as opposed to popping out of the display.

To correct for the negative parallax, we crop from the left and right views, in a way that object of interest appears on the screen, and rest of objects appear behind the screen. More details regarding this correction can be found in [19].

III. CHALLENGES, USE CASES, AND POTENTIAL RESEARCH OPPORTUNITIES IN PROCESSING 3D-HDR VIDEO

After post-processing, our SHDR video database is ready for use. That being said, there are multiple challenges with 3D-HDR data and requires special considerations. This section iterates some of these challenges.

*A. Tone Mapping and Dynamic Range Compression*

Tone mapping of HDR image content is a challenge, and still a very hot area of interest to research community. In the case of HDR video, it would be even more challenging as temporal coherency of tone-mapped video is crucial. Fig. 4 shows two difference scenes when they are over-exposed, under-exposed, and tone-mapped. It is observed from this figure, that "clouds" or "tree colors" can be hidden in the first two cases for the top row of this figure. In the bottom row of the figure, the "airplane" or building details can get lost in the two extremes of exposure. This figure shows the importance of an accurate tone-mapping technique.

In the case of 3D-HDR video, tone-mapping will be even more challenging, as not only temporal coherency is essential, but also inter-view coherency has to be assured. While human eyes can tolerate a minimal degree of difference between the tones of the two views due to binocular suppression, drastic tone differences between the views can severely degrade the overall video quality.

*B. 3D Specific Challenges*

Depth map extraction can potentially be a different task for stereoscopic HDR video compared to that of stereo SDR video. Most of the existing state-of-the-art depth synthesis methods are designed or tested against only SDR video. Therefore, depth estimation from stereo HDR video has room for further investigations. There will be questions such as: Do we need to tone map the SHDR video first, and then generate depth maps? Or do we generate depth maps from floating point values stored in HDR files? How should one consider dynamic range locally or globally when estimating a depth map? Fig. 5 shows sample depth maps from the presented stereo HDR videos. For demonstration purposes, depth maps are generated by tone-mapping frames using the MPEG Depth Estimation Reference Software (DERS) [20, 46].

Another challenge with 3D-HDR video is handling crosstalk (of 3D aspect) and ghosting (of HDR aspect). In 3D video, sometimes objects appear as double shadowed, that is called crosstalk, and is mainly resulted from the leakage of one eye view to the other eye due to imperfections of the 3D display or glasses. In HDR video, on the other hand, fast moving objects may appear blurry, because each frame is resulted from combining over a dozen F-stops within which a moving object may have different positions. In 3D-HDR domain, reducing the crosstalk and ghosting together will likely be a much more difficult task as they may both target a same object.

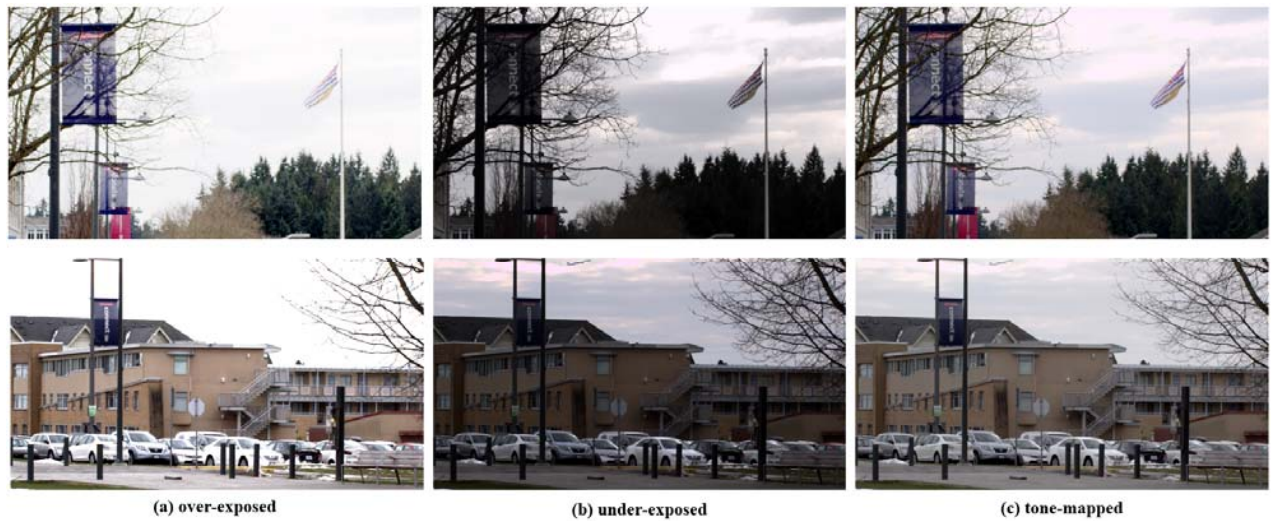

Fig. 4. Capturing with different exposure levels: (a) over-exposed, (b) under-exposed, and (c) tone-mapped. Note that tree colors are not visible in (b)-top, and airplane is not visible in (a)-bottom. Only right view frame is shown here.

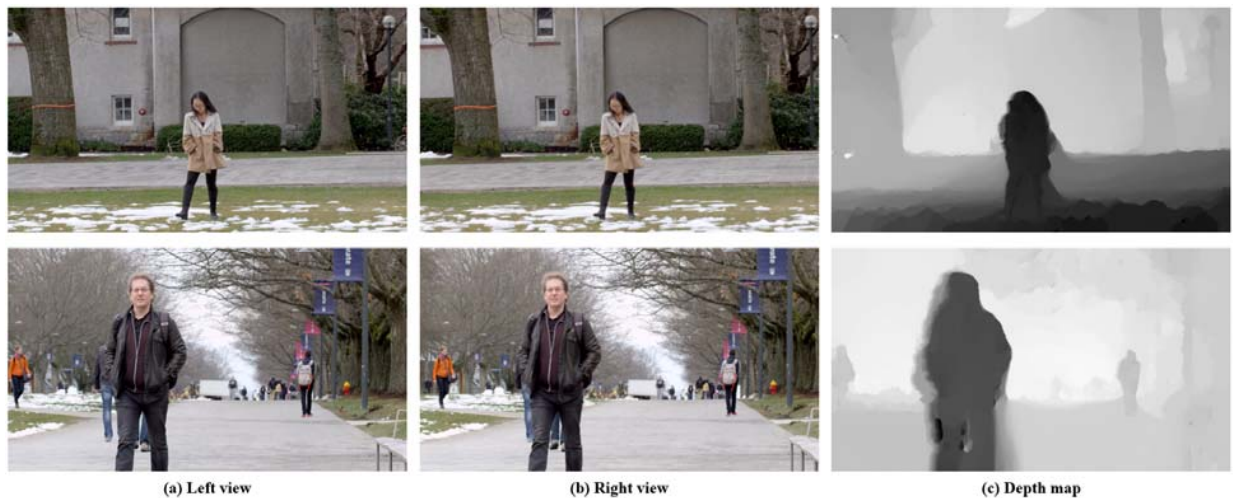

Fig. 5. Synthesized depth map samples.

*C. Other Challenges*

One other major challenge in the processing of 3D-HDR video is encoding such type of video content. MPEG/ITU are developing video compression standards for 3D or HDR video, but standards for compression of 3D-HDR video are yet to come.

Quality assessment of 3D-HDR video also opens new doors for potential introduction of new quality metrics that are dynamic range independent [15], and also consider binocular properties of human vision [5-8].

Last but not the least, visual attention modeling in 3D-HDR video needs to be studied, as it will be different from that of either 3D [9-10] or HDR [15-16] video.

## IV. CONCLUSION

With the research and industry communities showing an increasing amount of interest in advanced immersive media technologies, it is essential to have publicly available database to facilitate studying such media systems. This paper introduces a Stereoscopic 3D HDR video database made of 11 stereo HDR sequences from scenes with different characteristics. The capturing configuration and database characteristics are described first. Then post-processing stages were explained in details. It is important to provide reproducible steps so that interested readers can carry out the same approach to create similar databases from different scenes or under different capturing settings. At the end, some challenges and potential future research opportunities were discussed. This database is made publicly available to the research community.

FUNDING

This research did not receive any specific grant from funding agencies in the public, commercial, or not-for-profit sectors.